\documentclass[prd, preprint, showpacs, showkeys]{revtex4-2}
\usepackage{etoolbox}
\AfterEndPreamble{
    \setlength{\parindent}{1.5em}
}
\usepackage{indentfirst}
\usepackage{amsmath,amssymb}
\usepackage{graphicx}
\usepackage{bm}
\usepackage{ulem,soul} 
\usepackage{verbatim} 

\usepackage[utf8]{inputenc}
\usepackage{subcaption}
\usepackage{multirow}
\usepackage{array}
\usepackage{float}
\usepackage[compat=1.1.0]{tikz-feynman}
\usepackage{tikz}
\usepackage{slashed}
\usetikzlibrary{calc,arrows.meta}

\usepackage{xurl}  
\usepackage{hyperref}
\hypersetup{breaklinks=true}

\begin{document}

\title{Study the property of $W^{\prime}$ at future $e^-p$ collider}
\author{Xinyi Yan}
\author{Honglei Li}\email{sps\_lihl@ujn.edu.cn}
\author{Zhi-Long Han}\email{sps\_hanzl@ujn.edu.cn}
\author{Fei Huang}\email{sps\_huangf@ujn.edu.cn}
\author{Ruiyu Xing}
\affiliation{School of Physics and Technology, \\ 
University of Jinan, 250022, Jinan, China}

\begin{abstract}
As a strong candidate for new physics beyond the Standard Model, the exotic charged gauge boson $W^{\prime}$ has attracted extensive research interest. In this work we investigate the interactions of the $W^{\prime}$ boson  at the electron-proton colliders.  The process $e^- u \to \nu_e d$ and $e^- u \to e^\pm jjj$ with $t$-channel $W^{\prime}$ exchange are studied. The polarization of the initial-state electrons has a significant impact on the cross section of the studied process, while the angular distribution of the final-state leptons serves as an important observable for the interactions of the $W^{\prime}$ boson. In some specific regions of the parameter space, the detectable mass range for the $W^{\prime}$ boson can reach around 10 TeV, and the coupling strength can achieve a precision of approximately 1\% relative to the interaction strength of the Standard Model. Especially, $e^- u \to e^+ jjj$ process is forbidden within the Standard Model, which would constitute important evidence in the search for the Left-Right Symmetric Model.
\end{abstract}
\maketitle
\newpage
\quad

\section{Introduction}
The Standard Model (SM)\cite{Thomson:2013zua,Gaillard:1998ui,CDF:2022hxs} of particle physics is a theoretical framework formulated in the language of quantum field theory, describing the behavior of known fundamental particles under the strong, weak, and electromagnetic interactions. It has proven to be a powerful tool for studying particle physics. However, with the advancement of theoretical and experimental research in particle physics, phenomena such as dark matter\cite{Baudis:2025yva,Essig:2013lka}, dark energy\cite{Dodelson:1993je,Lahav:2024npe}, and neutrino oscillations\cite{Super-Kamiokande:1998kpq} have been discovered, which cannot be explained within the SM. 
Motivated by these issues, physicists have proposed various extensions of the SM, leading to a wide range of new physics theories and models. In recent years, researches on these new physics models have achieved remarkable success, and many of them are expected to be tested in current and future experiments.

One of the most compelling new physics models is to extend the gauge sector of the SM. Many such models predict the existence of heavy charged vector bosons, denoted as $W^{\prime}$. Examples include Left-Right Symmetric Models, Little Higgs Models, and Extra Dimension Models\cite{Polesello:2003ch,YaserAyazi:2010oym,Barman:2022qix}. Among these, the Left-Right Symmetric Model  demonstrates notable advantages in addressing the CP problem, neutrino masses, and other related phenomena . It introduces an additional $SU(2)$ gauge group beyond the SM, one degree of freedom of which gives rise to a heavy gauge boson $W^{\prime}$. Extensive phenomenological studies related to the $W^{\prime}$ boson have been conducted successively.

In recent years, several experimental collaborations have systematically imposed stringent limits on the mass of the $W^{\prime}$ boson using 13 TeV proton–proton collision data from the LHC. Under the Sequential Standard Model (SSM) with the assumption that the right-handed couplings are equal to the Standard Model couplings, the CMS Collaboration excluded $M_{W^{\prime}} < 4.0\;\text{TeV}$ via the $\tau\nu$ decay channel with an integrated luminosity of $35.9\;\text{fb}^{-1}$, and raised this lower limit to $5.3\;\text{TeV}$ in the electron decay channel using the full Run-2 dataset. For the Left-Right Symmetric Model (LRSM) with the assumption that the right-handed neutrino mass is half the $W^{\prime}$ mass, exclusion limits reach $4.7\;\text{TeV}$ and $5.0\;\text{TeV}$ in the electron and muon channels, respectively\cite{CMS:2018fza,CMS:2022krd,CMS:2021dzb}. Utilizing $138\;\text{fb}^{-1}$ of data, the ATLAS Collaboration pushed the mass limit to $M_{W^{\prime}} > 5.0\;\text{TeV}$ in the $\tau\nu$ channel within the SSM framework. When considering non-universal couplings ($g_{L} \neq g_{R}$), the $\tau\nu$ channel further excludes the mass range $3.5–5.0\;\text{TeV}$. Under the same LRSM hypothesis, exclusion limits of $4.8\;\text{TeV}$ and $5.0\;\text{TeV}$ are obtained in the electron and muon channels, respectively\cite{ATLAS:2024tzc,ATLAS:2019isd}. 
The electron-proton ($e^-p$) collider demonstrates significant advantages in studying nucleon structure, parton distribution functions, strong interactions, and the search for new physics\cite{LHeC:2020van,LHeCStudyGroup:2012zhm,Antusch:2020fyz,Bruening:2013bga,FCC:2018byv,Spor:2021dhf,Gutierrez-Rodriguez:2020gsi}.The European Organization for Nuclear Research (CERN) has designed two major future colliders for deep inelastic lepton–hadron scattering: the Large Hadron Electron Collider (LHeC)\cite{LHeCStudyGroup:2012zhm,LHeC:2020van,Antusch:2020fyz,Bruening:2013bga} and the Future Circular Collider–Hadron Electron (FCC-he)\cite{Spor:2021dhf,Gutierrez-Rodriguez:2020gsi,Spor:2021dhf}, both aiming to integrate the proton beams of the Large Hadron Collider (LHC) with new electron accelerators in the LHC main tunnel. The center-of-mass energy $\sqrt{s}$ and integrated luminosity $\mathcal{L}$ of the LHeC and FCC-he are summarized in Table \ref{tab:1}. Based on the relevant experimental designs, we aim to investigate the properties of the $W^{\prime}$ boson at $e^-p$ colliders.

\begin{table}[htbp]
    \centering
    \renewcommand{\arraystretch}{1.8}
    \begin{tabular}{|c|c|c|c|c|}
         \hline 
        Colliders & $ \sqrt{s}$(TeV) & $E_e$(GeV) &$E_p$(TeV)&
$\mathcal{L}\left(\mathrm{fb}^{-1}\right)$\\ \hline 
   \multirow{2}{*}{LHeC}    & 1.30 & 60 & 7 &  \multirow{2}{*}{10,30,50,100}      \\ \cline{2-4}
                           & 1.98 & 140 & 7   & \\ \hline
    \multirow{2}{*}{FCC-he}    & 3.46 & 60 & 50   &  \multirow{2}{*}{100,500,1000,2000}      \\ \cline{2-4}
                           & 5.29 & 140 & 50    &       \\ \hline  

    \end{tabular}
    \caption{The collision energy  and integrated luminosity of the two types of \( e^-p \) colliders.}
    \label{tab:1}
\end{table}

This paper is organized as follows.
In Section~\ref{sec:2}, we present a detailed description of the theoretical framework.
In Section~\ref{sec:3}, we perform a comprehensive study of the scattering cross sections and angular distributions of the final-state particles, as well as the asymmetries induced by the different $W^{\prime}$ decay modes, for the processes $e^- u \to \nu_e d$ and $e^- u \to e^\pm jjj$. A brief summary is given in Section~\ref{sec:4}.

\section{THEORETICAL MODEL FRAMEWORK}\label{sec:2}
To illuminate the origin  of neutrino masses,  dark matter, and the cosmic baryon--anti-baryon asymmetry, the Standard Model is extended to an attractive and renormalisable framework---the Left--Right Symmetric Model (LRSM)\cite{Senjanovic:1975rk,Mohapatra:1974gc,Pati:1974yy,Zhao:2024fgc,Lu:2024qrf} in which both type-I and type-II seesaw mechanisms can be naturally embedded.
The model is based on the gauge group
\begin{equation}
\mathrm{SU}(3)_C\otimes\mathrm{SU}(2)_L\otimes\mathrm{SU}(2)_R\otimes\mathrm{U}(1)_{B-L},
\end{equation}
and necessarily predicts right-handed (RH) charged currents together with heavy RH gauge bosons $W^{\prime}$ and $Z^{\prime}$.
In addition, the LRSM contains three RH neutrinos $N$ which are charged under $\mathrm{SU}(2)_R\otimes\mathrm{U}(1)_{B-L}$ but remain singlets of the SM gauge symmetry. For symmetry-breaking scales of a few TeV, such a effective LRSM can be directly tested at the Large Hadron Collider (LHC) through the resonant production modes\cite{Fuks:2017vtl,Huang:2017yso,Mattelaer:2016ynf,Mitra:2016kov,Gong:2014qla,Frederix:2021zsh,Li:2015kaa,Bao:2011nh,CMS:2024xdq}
\begin{itemize}
  \item $pp\to W^{\prime \pm}\to tb$\, or\, $N\ell^\pm$ \; with \; $N\to\ell^\pm W^{\prime \mp }\to\ell^\pm jj$,
  \item $pp\to Z^{\prime} \to t\bar t$\, or\, $NN$.
\end{itemize}
These channels constrain the masses and couplings of $W^{\prime}$, $Z^{\prime}$ and $N$, and give rise to a variety of experimental signatures such as same-sign dileptons plus jets ($\ell^\pm\ell^\pm+nj$)\cite{Keung:1983uu}, single-lepton plus jets ($\ell^\pm+j$)\cite{Mitra:2016kov}, and single-top\cite{Simmons:1996ws} final states, all of which have been extensively studied by the ATLAS and CMS collaborations\cite{Han:2012vk,Frank:2010vc,Tello:2010am,Das:2012ii,Chen:2013qba,Vasquez:2016jqb,Maiezza:2015lza,Gluza:2015goa}.

The effective low-energy field content of the Effective Left--Right Symmetric Model (LRSM) comprises the usual Standard-Model states, the gauge bosons $W^{\prime}$ and $Z^{\prime}$ (aligned with their mass eigenstates), and three heavy Majorana neutrinos $N_i$ whose chiral components are purely right-handed.  In the LRSM the chiral couplings of $W^{\prime}$ to quarks are given by
\begin{equation}
\mathcal{L}_{W^{\prime}-q-q^{\prime}}=\frac{-\kappa_{R}^{q} g}{\sqrt{2}} \sum_{i, j=u, d, \ldots} \bar{u}_{i} V_{i j}^{\mathrm{CKM}^{\prime}} W^{\prime+}_{\mu} \gamma^{\mu} P_{R} d_{j}+\text{H.c.} ~.
\end{equation}

where $u_i$ ($d_j$) denotes the up-type (down-type) quark of flavour $i$ ($j$);
$P_{R(L)}=\frac12(1\pm\gamma_5)$ is the chiral projector for right-(left-)handed states;
$V_{ij}^{\text{CKM}'}$ is the right-handed Cabibbo--Kobayashi--Maskawa (CKM) matrix, related to the SM CKM matrix.
Throughout this study we work in the five massless-quark limit and take both the SM and the RH CKM matrices to be diagonal with unit entries.
Finally, $g=\sqrt{4\pi\alpha_{\text{EM}}(M_Z)}/\sin\theta_w$ is the SM weak coupling constant and $\kappa_{R}^q$ is an overall normalization the strength of the $W^{\prime}$ interaction, which is a coefficient related to the coupling of the $W^{\prime}$ boson to fermions.
For the leptons, the $W^{\prime}$ coupling and lepton mixing are parametrized as\cite{Atre:2009rg}
\begin{equation}
\mathcal{L}_{W^{\prime}- \ell -\nu / N}
= \frac{-\kappa^\ell_R\, g}{\sqrt{2}}
\sum_{\ell = e, \mu, \tau}
\Bigg[
\sum_{m=1}^{3} \overline{\nu^c_m}\, X_{\ell m}
+
\sum_{m^{\prime}=1}^{3} \overline{N_{m^{\prime}}}\, Y_{\ell m'}
\Bigg]
\,W^{^{\prime}+}_{\mu}\,\gamma^{\mu}\,P_R\,\ell^{-}
+\text{H.c.} ~.
\end{equation}
The matrix $Y_{\ell m^{\prime}}$ ($X_{\ell m}$) quantifies the mixing between the heavy (light) neutrino mass eigenstate $N_{m^{\prime}}$ ($\nu_{m}$) the RH chiral state with the corresponding lepton flavor $\ell$. The mixing ratio is  \cite{Keung:1983uu}
\begin{equation}
|Y_{\ell m'}|^2 \sim \mathcal{O}(1) \quad \text{and} \quad 
|X_{\ell m}|^2 \sim 1 - |Y_{\ell m'}|^2 \sim \mathcal{O}\!\left(\frac{m_{\nu_m}}{m_{N_{m'}}}\right).
\end{equation}

As in the quark sector, $\kappa^{\ell}_R\in\mathbb{R}$ independently normalises the $W^{\prime}$ coupling strength to the leptons.  
At TeV-scale collider energies both the light-neutrino masses and their mixings can be taken to vanish.  
For simplicity we therefore set the matrix $Y_{\ell m'}$ to a diagonal unit matrix:
\begin{equation}
|Y_{eN}| = |Y_{\mu N_2}| = |Y_{\tau N_3}| = 1,\quad |Y_{\text{others}}| = |X_{\ell m}| = 0.
\end{equation}

After LR symmetry breaking, the $W^{\prime 3}$ and $X_{(B-L)}$ gauge states mix and produce the large mass $Z^{\prime}$ and massless (hypercharge) $B$ bosons. Subsequently, all the fermions with $(B-L)$ charge, including $\nu_{L}$ and $N$, couple to the $Z^{\prime}$.The electromagnetic quantum numbers of the fermion field $f$ under $\mathrm{SU}(2)_L$, $\mathrm{SU}(2)_R$ and $\mathrm{U}(1)$ are summarised in Table \ref{tab:my-table}.

\begin{table}[h]
	\centering
 \renewcommand{\arraystretch}{1.8}
	\begin{tabular}{|l|l|l|l|l|l|l|l|l|l|}
	\hline
		 Gauge Group & Charge &
        $\mathbf{u_{L}}$ & $\mathbf{d_{L}}$ &
        $\boldsymbol{\nu_{L}}$ & $\mathbf{e_{L}}$ &
        $\mathbf{u_{R}}$ & $\mathbf{d_{R}}$ &
        $\mathbf{N_{R}}$ & $\mathbf{e_{R}}$ \\ \hline\hline
		$SU(2)_{L}$ &$T_{L}^{3,f}$ &
        $+\frac{1}{2}$ & $-\frac{1}{2}$ &
        $+\frac{1}{2}$ & $-\frac{1}{2}$ &
        $0$ & $0$ & $0$ & $0$ \\ \hline
		$SU(2)_{R}$ &
        $T_{R}^{3,f}$ &
        $0$ & $0$ &
        $0$ & $0$ &
        $+\frac{1}{2}$ & $-\frac{1}{2}$ &
        $+\frac{1}{2}$ & $-\frac{1}{2}$ \\ \hline
		$U(1)_{\mathrm{EM}}$ &
        $Q^{f}$ &
        $+\frac{2}{3}$ & $-\frac{1}{3}$ &
        $0$ & $-1$ &
        $+\frac{2}{3}$ & $-\frac{1}{3}$ &
        $0$ & $-1$ \\ \hline
	\end{tabular}
 \captionsetup{justification=raggedright}
	\caption{$\mathrm{SU}(2)_L$, $\mathrm{SU}(2)_R$, and $\mathrm{U}(1)_\mathrm{EM}$ quantum number assignments for chiral fermions $f$ in LRSM.}
	\label{tab:my-table}
\end{table}

For generic $\kappa^{q,\ell}_{R}$ normalizations, the leading order $W^{\prime}$ partial decay widths are then   

\begin{equation}
\Gamma\left(W^{\prime} \rightarrow q \overline{q^{\prime}}\right) = N_{c}\left|V_{q q^{\prime}}^{\mathrm{CKM}^{\prime}}\right|^{2} \frac{\kappa_{R}^{q 2} g^{2} M_{W^{\prime}}}{48 \pi}
\end{equation}
\begin{equation}
\Gamma\left(W^{\prime} \rightarrow t b\right) = N_{c}\left|V_{t b}^{\mathrm{CKM}^{\prime}}\right|^{2} \frac{\kappa_{R}^{q 2} g^{2} M_{W^{\prime}}}{48 \pi}\left(1-r_{t}^{W^{\prime}}\right)^{2}\left(1+\frac{1}{2} r_{t}^{W^{\prime}}\right)
\end{equation}
\begin{equation}
\Gamma\left(W^{\prime} \rightarrow \ell N_{m^{\prime}}\right) = \left|Y_{\ell N_{m^{\prime}}}\right|^{2} \frac{\kappa_{R}^{\ell 2} g^{2} M_{W^{\prime}}}{48 \pi}\left(1-r_{N}^{W^{\prime}}\right)^{2}\left(1+\frac{1}{2} r_{N}^{W^{\prime}}\right)
\end{equation}

with the mass ratio defined as $r_i^{W^{\prime}} \equiv m_i^2 / {M_{W^{\prime}}}^2$, where $m_i$ is the mass of the final-state fermion.

In other new physics models based on the Left-Right Symmetry Model, extensions such as the Higgs sector are also included. This study focuses on the production and interactions of the $W^{\prime}$ boson at $e^-p$ collider, for which the Effective Left-Right Symmetric Model already provides a sufficient theoretical framework. 

\section{Production and Decay of $W'$ Boson at Electron-Proton Colliders}\label{sec:3}
In this section, we will systematically  investigate  the production mechanisms and decay properties of heavy gauge boson $W^\prime$ at future electron-proton colliders. We first analyze the  process $e^- u \rightarrow \nu_e d $ within the framework of a simple model --- the $W^\prime$ Effective Model\cite{Sullivan:2002jt,Duffty:2012rf}, evaluating its detection sensitivity and theoretical constraints. In this model, only the $W^{\prime}$ boson is introduced, with no extensions to other particles or fermion fields. 
Subsequently, within an effective field theory framework based on the symmetry breaking, we focus on the multi-body final-state process $e^- u \rightarrow e^\pm j j j$ in the Effective Left–Right Symmetric Model, which involves lepton number violation signals mediated by $t$-channel $W^\prime$ exchange and Majorana neutrinos $N$. Through systematic calculations of differential cross sections, angular distributions, forward-backward asymmetries, and statistical significance, we establish feasible windows for probing new physics at the TeV scale in the electron-proton collider environment.

\subsection{The process $e^- u \to \nu_e d$ in $W^{\prime}$ Effective Model}
\begin{figure}[]
	\centering
\includegraphics[width=0.8\linewidth]{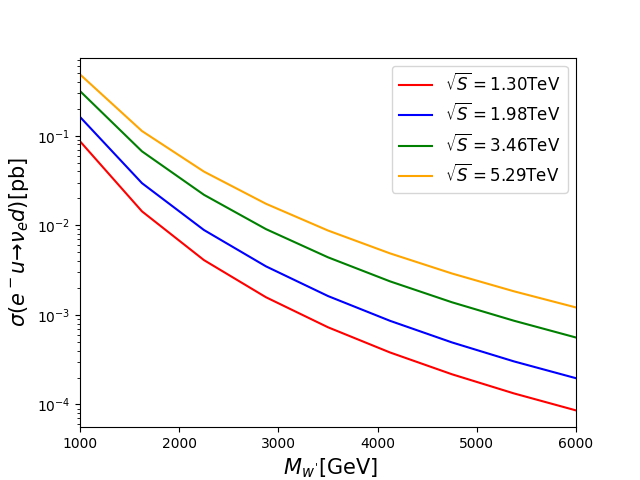}
\captionsetup{justification=raggedright}
	\caption{In the $W^{\prime}$ Effective Model, the distribution of cross section varies with $M_{W^{\prime}}$ based on the energy requirements of electrons and protons in LHeC and FCC-he.}
 \label{fig:cross_section}
\end{figure}

In electron-proton collisions, the process $e^- u \rightarrow \nu_e d$ can proceed via $t$-channel exchange of a charged gauge boson $W^{\prime}$.
Based on the design center-of-mass energies of current electron-proton colliders (as listed in Table~\ref{tab:1}), Fig.~\ref{fig:cross_section} presents the cross section for the process $e^- u \rightarrow \nu_e d$ as a function of the $W^\prime$ boson mass ($M_{W^\prime}$) at various $\sqrt{s}$ values for the LHeC and FCC-he. For LHeC, the proton beam energy is fixed at 7 TeV, with electron beam energies of 60 GeV and 140 GeV. For FCC-he , the proton beam energy is fixed at 50 TeV, with electron energies also at 60 GeV and 140 GeV.The results show that, for all considered center-of-mass energies, the cross section decreases monotonically with increasing $M_{W^\prime}$, consistent with the behavior expected from the $t$-channel propagator in lepton-parton scattering. A higher electron beam energy corresponds to a larger $\sqrt{s}$, leading to a larger cross section. In the mass range $M_{W^\prime} = 1 \sim 5$ TeV, the cross section varies approximately between $ 10^{-1} \sim 10^{-4}$ pb. Comparing the cross section behavior at different energies provides guidance for experimental energy selection to optimize signal detection conditions. However, when $M_{W^\prime}$ is approximately $1~\text{TeV}$, the cross section is only about 1 pb, resulting in a relatively low event incidence rate and bringing significant observational challenges.

\begin{figure}[]
	\centering
	\subfloat[]{
		\includegraphics[width=0.48\linewidth]{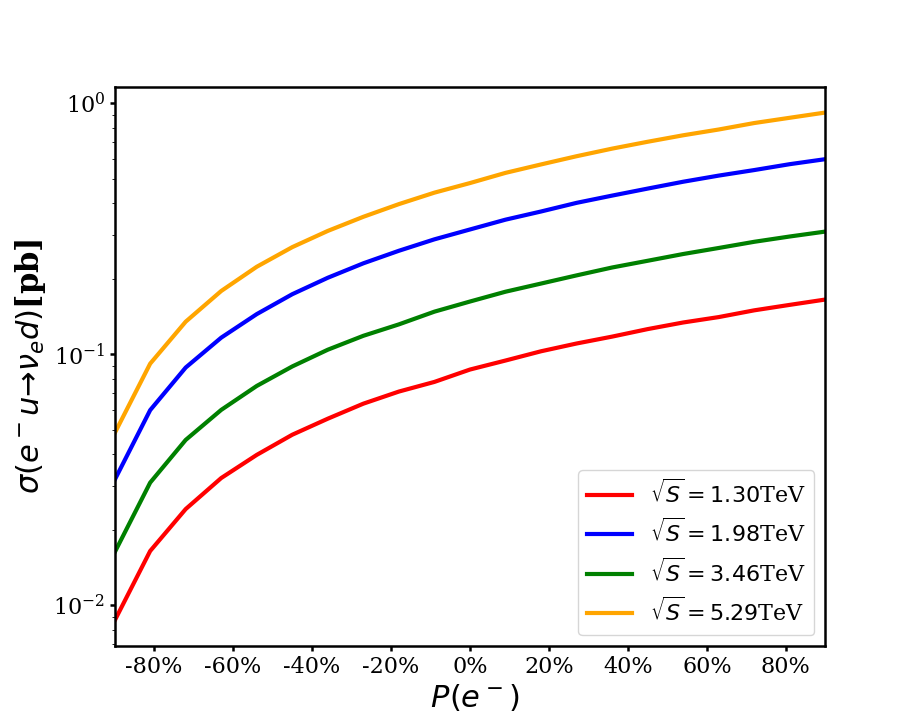}
	} \hfill
	\subfloat[]{
		\includegraphics[width=0.48\linewidth]{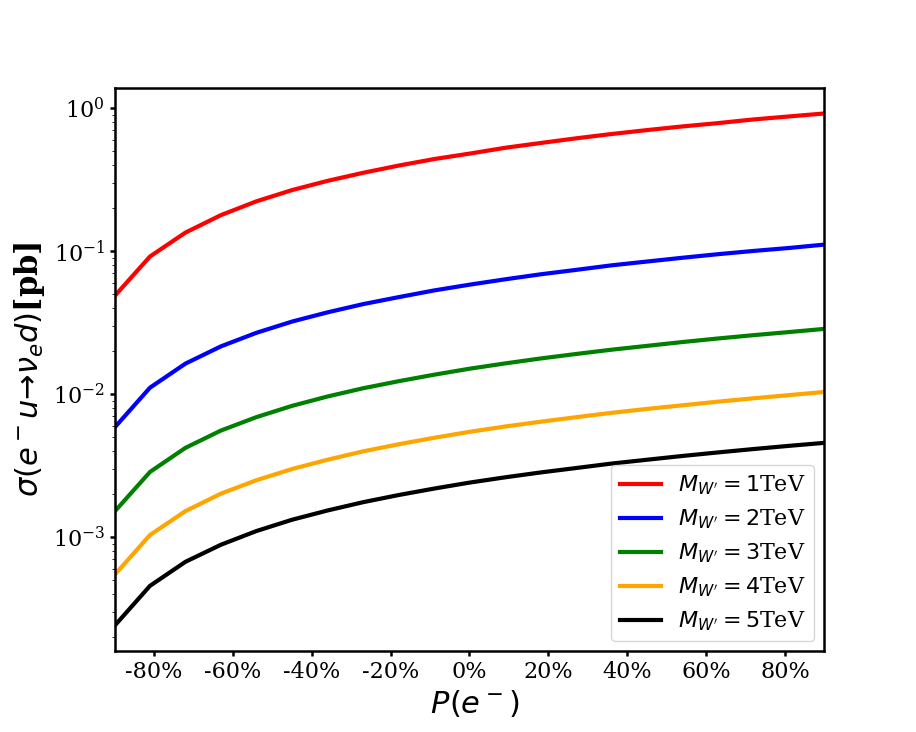}
	}
\captionsetup{justification=raggedright}
	\caption{ The distribution of the cross section varying with polarization of initial-state electrons in the process $e^- u \rightarrow \nu_e d$. (a) $M_{W'}=1~\text{TeV}$. (b)   $\sqrt{s} = 5.29~\text{TeV}$.}
	\label{P4}
\end{figure}

Polarization serves as a critical probe in particle physics experiments, allowing the enhancement of specific interaction channels by controlling the spin orientation of initial-state particles. This technique significantly increases the sensitivity to coupling structures and new physics energy scales. Fig.~\ref{P4}  illustrates the dependence of the $e^- u \rightarrow \nu_e d$ scattering cross section on the electron beam polarization degree $P(e^-)$. For example,  $P(e^-)=80\% (-80\%)$ means that 80\% of the initial-state electrons are right(left)-handed polarized.The  cross section is calculated with different center-of-mass energies of $1.30, 1.98, 3.46$,  and $5.29~\text{TeV}$. In Fig.~\ref{P4}(a), it can be observed that at all energy points, the cross section increases monotonically with $P(e^-)$, indicating a clear dominance of right-handed current couplings in the process. Furthermore, the cross section rises significantly with increasing collision energy. In Fig.~\ref{P4}(b) the collision energy is fixed at $\sqrt{s} = 5.29~\text{TeV}$ and the  $W^{\prime}$ boson masses are set at some typical points in the range of $1 \sim 5~\text{TeV}$.  The results show that at the same polarization degree, the cross section decreases with increasing $M_{W^{\prime}}$, consistent with the mass suppression effect from the $t$-channel propagator. In particular, for $M_{W^{\prime}} = 1~\text{TeV}$ and $P(e^-)=80\%$, the cross section is 0.87 pb. As the electron polarization approaches complete positive polarization ($P(e^-) \rightarrow 1$), all curves approach their respective asymptotic maxima, further confirming the dominance of right-handed electron current couplings in this process. The above analysis indicates that combining high-energy collisions with a highly polarized electron beam can effectively enhance the sensitivity to potential $W^{\prime}$ boson signals or other new physics phenomena. 

\begin{figure}[]
	\centering
	\subfloat[]{
		\includegraphics[width=0.48\linewidth]{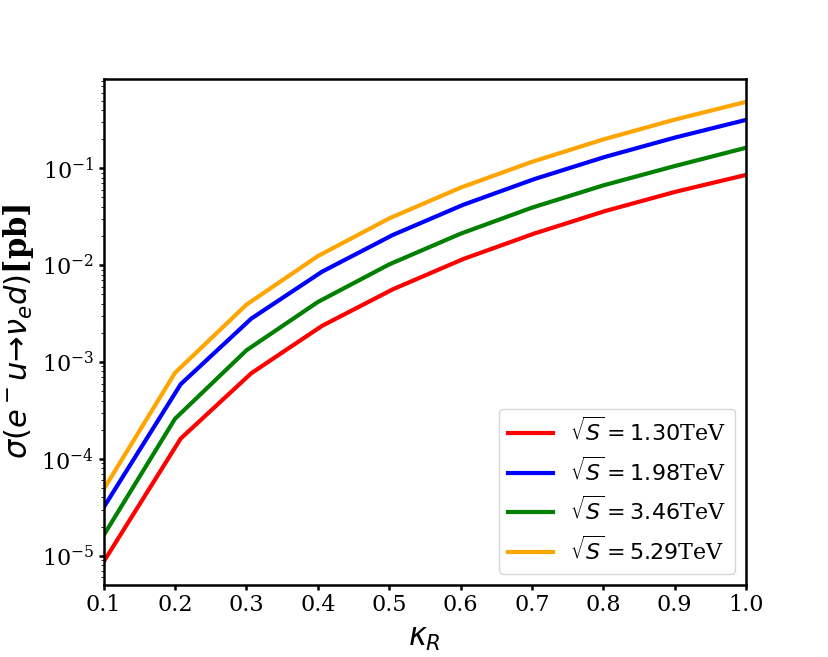}
	} \hfill
	\subfloat[]{
		\includegraphics[width=0.48\linewidth]{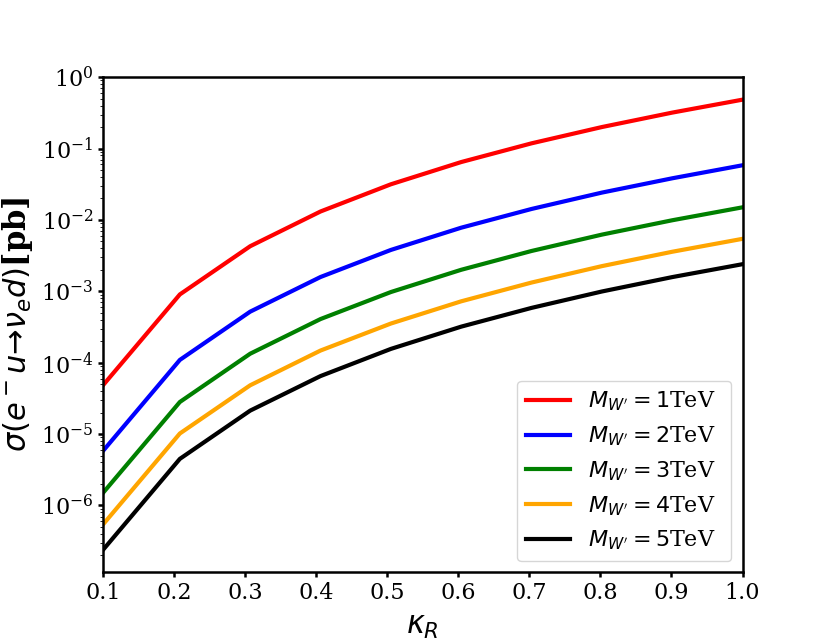}
	}
 \captionsetup{justification=raggedright}
	\caption{The distribution of cross section varies with the increase of $\kappa_R $. (a) $M_{W^{\prime}} $=1 TeV. (b) $\sqrt{s} = 5.29~\text{TeV}$. }
	\label{kr4}
\end{figure}

 In the $W^{\prime}$ Effective Model, the coupling strength between $W^{\prime}$ and leptons is determined by the right-handed coupling parameter $\kappa_R$. Fig.~\ref{kr4}(a) displays the variation of the cross section with $\kappa_R$ for a fixed $W^{\prime}$ boson mass $m_{W^{\prime}} = 1\ \text{TeV}$. Different colors of the curves represent different center-of-mass energies $\sqrt{s} = 1.30$, 1.98, 3.46, and 5.29\ \text{TeV}. The cross section increases with the growth of $\kappa_R$, and this rising trend exhibits a consistent pattern across different collision energies. At the center-of-mass energy $\sqrt{s} = 5.29\ \text{TeV}$ and $\kappa_R = 1$, the cross section reaches $0.48\ \text{pb}$. In Fig.~\ref{kr4}(b), with the center-of-mass energy fixed at $\sqrt{s} = 5.29\ \text{TeV}$, the cross section is shown as a function of $\kappa_R$ for several $W^{\prime}$ boson masses in the range of $M_{W^{\prime}} = 1\sim 5\ \text{TeV}$. For a fixed $M_{W^{\prime}}$, the cross section rises with increasing $\kappa_R$. For $M_{W^{\prime}} = 3\ \text{TeV}$ and $\kappa_R=0.8$, the cross section is $6.228\times10^{-3}$ pb.

\subsection{The process $e^{-}u\to e^{\pm}jjj$ in The Effective Left–Right Symmetric Model}
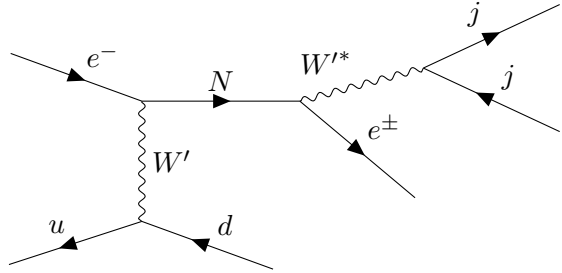
\begin{figure}[htbp]
\centering
	\begin{tikzpicture}
  \begin{feynman}
    \vertex (v1);
    \vertex [below=1.6cm of v1] (v2);
   \vertex (i1) at ([shift=(160:2cm)] v1) {};
\vertex (i2) at ([shift=(198:2cm)] v2) {};
    \vertex [right=2.1cm of v1] (f1);
    \vertex (f2) at ([shift=(-20:2cm)] v2) {};
    \vertex [right=0cm of f1] (tauDecay);
    \vertex at ($(tauDecay) + (15:1.7cm)$) (Zprime);
    \vertex at ($(tauDecay) + (-40:2cm)$) (mu1);
    \vertex [right=0cm of Zprime] (ZprimeDecay);
    \vertex at ($(ZprimeDecay) + (25:2cm)$) (tauFinal);
    \vertex at ($(ZprimeDecay) + (-25:2cm)$) (mu2);
    \diagram* {
      (i1) -- [fermion, edge label=\(e^-\)] (v1),
      (i2) -- [anti fermion, edge label=\(u\)] (v2),
      (v1) -- [boson, edge label=\(W^{\prime}\), vertical] (v2),
      (v1) -- [fermion, edge label=\(N\)] (f1),
      (v2) -- [anti fermion, edge label=\(d\)] (f2),
      (f1) -- [fermion] (tauDecay),
      (tauDecay) -- [boson, edge label=\({W^{\prime}}^*\)] (Zprime),
      (tauDecay) -- [fermion, edge label=\(e^\pm\)] (mu1),
      (Zprime) -- [boson] (ZprimeDecay),
      (ZprimeDecay) -- [fermion, edge label=$j$] (tauFinal),
      (ZprimeDecay) -- [anti fermion, edge label=$j$] (mu2)
    };
  \end{feynman}
\end{tikzpicture}
\captionsetup{justification=raggedright}
\caption{The Feynman diagram of the $e^{-}u\to N d \to e^{\pm}jjj$ process.}
\label{fig:feynman}
\end{figure}

In the preceding section, we investigate the $ e^- u \to \nu_e d $ process within a simple model containing a $ W^{\prime}$ boson, exploring its detection potential. Subsequently, we turn to a more realistic framework—the Effective Left-Right Symmetric Model. In this scenario, the  $e^-u \to \nu_e d$  channel, which proceeds via  $t$-channel $W^{\prime}$ exchange, is directly forbidden for the left-right symmetry. Therefore, we focus  on the processes with charged leptons and multi-jet final states, such as $ e^-u \to Nd \to e^\pm jjj $, Which exhibit more distinctive signal features with $W^{\prime}$ intermediate. More importantly, the corresponding positron production process involves lepton number violation, which is of great significance for our research on new physics beyond the Standard Model. A representative Feynman diagram is shown in Fig.~\ref{fig:feynman}, involving $t$-channel $W'$ exchange and potentially including processes mediated by right-handed neutrino $N$.

\begin{figure}[htb]
	\centering
	\subfloat[]{
		\includegraphics[width=0.48\linewidth]{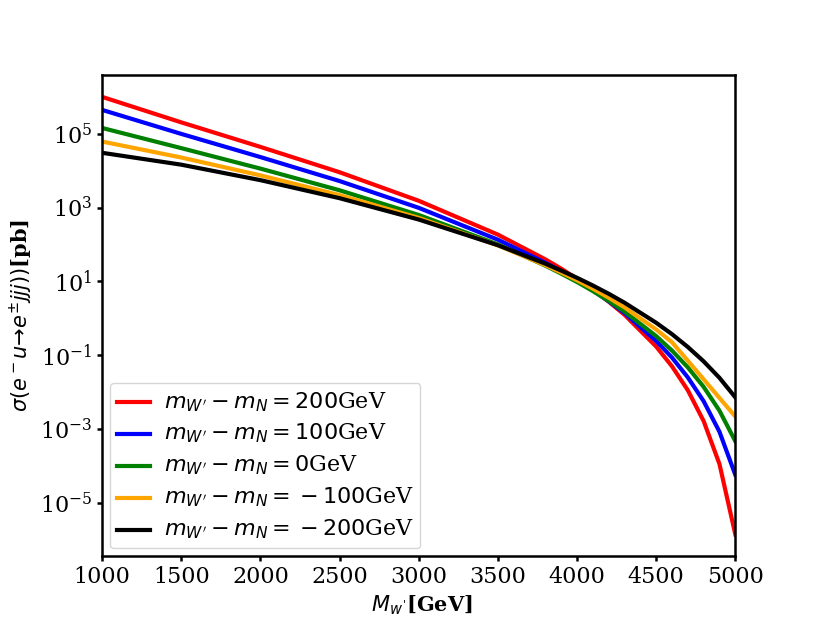}
	} 
	\subfloat[]{
		\includegraphics[width=0.48\linewidth]{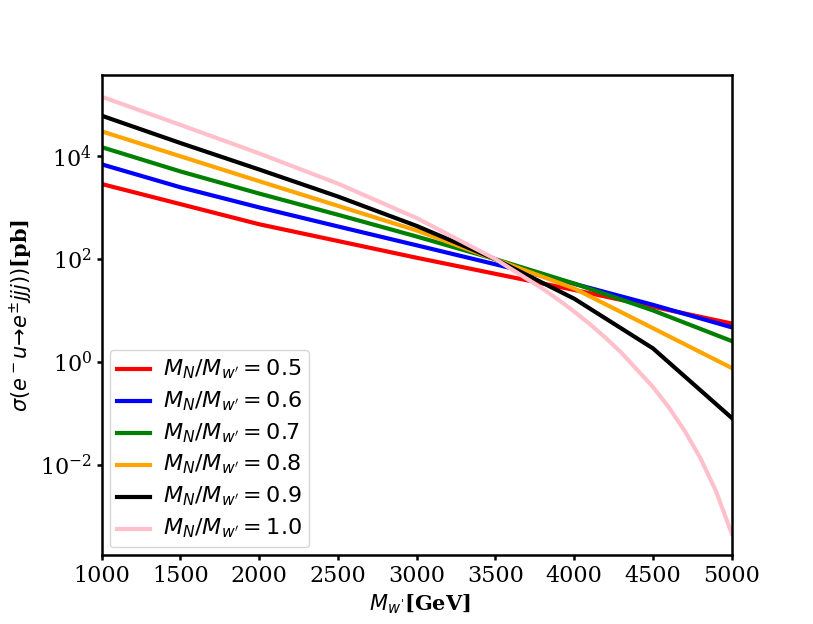}
	}
\captionsetup{justification=raggedright}
	\caption{The distribution of cross section  varies with different values of $M_{W^\prime}$ and $M_N$ in the process of $e^ -u \rightarrow e^\pm j j j$ with $\sqrt{s}=5.29~\text{TeV}$. }
   \label{fig:multijet_cross_section}
\end{figure}

The magnitude of the cross section for the process of $e^- u \rightarrow e^\pm j j j$ is influenced by the masses of the $W'$ and $N$ particles. In Fig.~\ref{fig:multijet_cross_section}, we adopt different parameter settings to investigate this process with the center-of-mass energy $\sqrt{s} = 5.29$ TeV (the design energy for FCC-he). In Fig.~\ref{fig:multijet_cross_section}(a), for the fixed value of mass splitting $\Delta M = M_{W^\prime} - M_N = 200, 100, 0, -100, -200$ GeV, the distribution of cross section is expressed as a function of $M_{W^\prime}$. The results show that it as $M_{W^\prime}$ increases, the cross section rapidly decreases. Considering different mass splitting scenarios, when the mass of $W^\prime$  is below 4 TeV, the cross section is relatively larger when the $W^\prime$ mass is greater than the $N$ mass. However, for $W^\prime$ masses above 4 TeV, the situation is reversed. Additionally, the intersection points of the curves in the figure will shift towards higher masses as the collision energy increases. In Fig.~\ref{fig:multijet_cross_section}(b) , with a fixed mass ratio of $M_N/M_{W^\prime}$ ranging from 0.5 to 1.0, the distribution of cross section also decreases as $M_{W^\prime}$ increases. As the mass ratio approaches 1, the cross section exhibits a more rapid decline with increasing mass. These results indicate that the cross section of $e^- u \rightarrow e^\pm j j j$ can be significantly enhanced under specific mass-splitting parameter conditions between $W^\prime$ and $N$.

\begin{figure}[htb]
	\centering
	\subfloat[]{
		\includegraphics[width=0.48\linewidth]{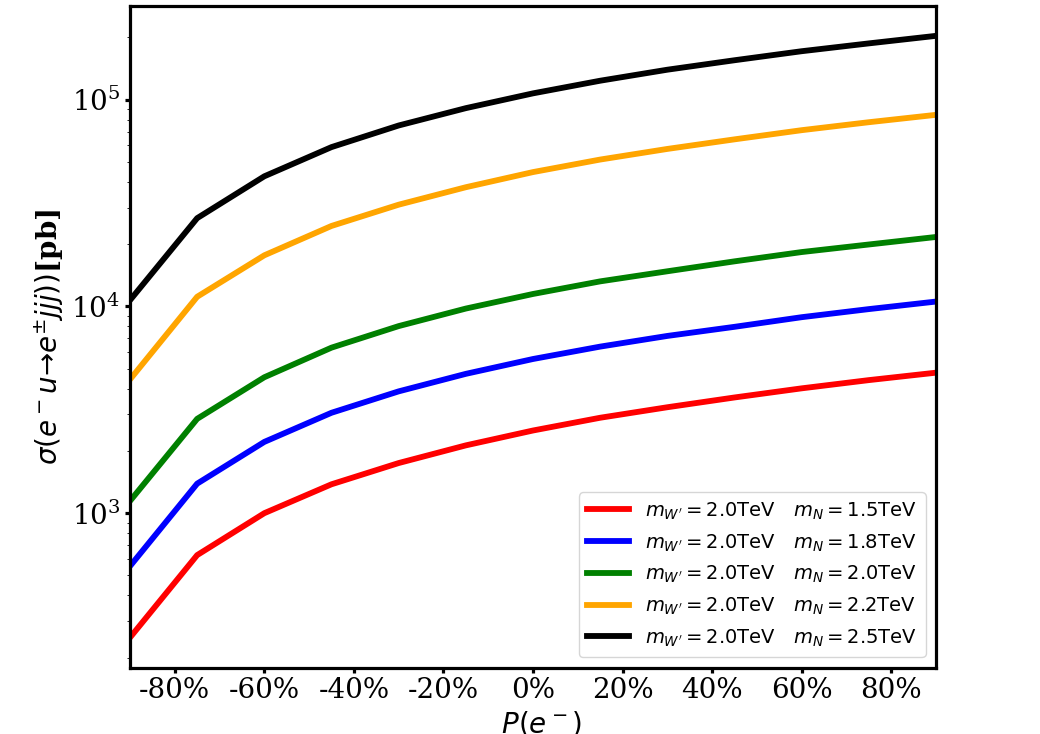}
	} 
	\subfloat[]{
		\includegraphics[width=0.48\linewidth]{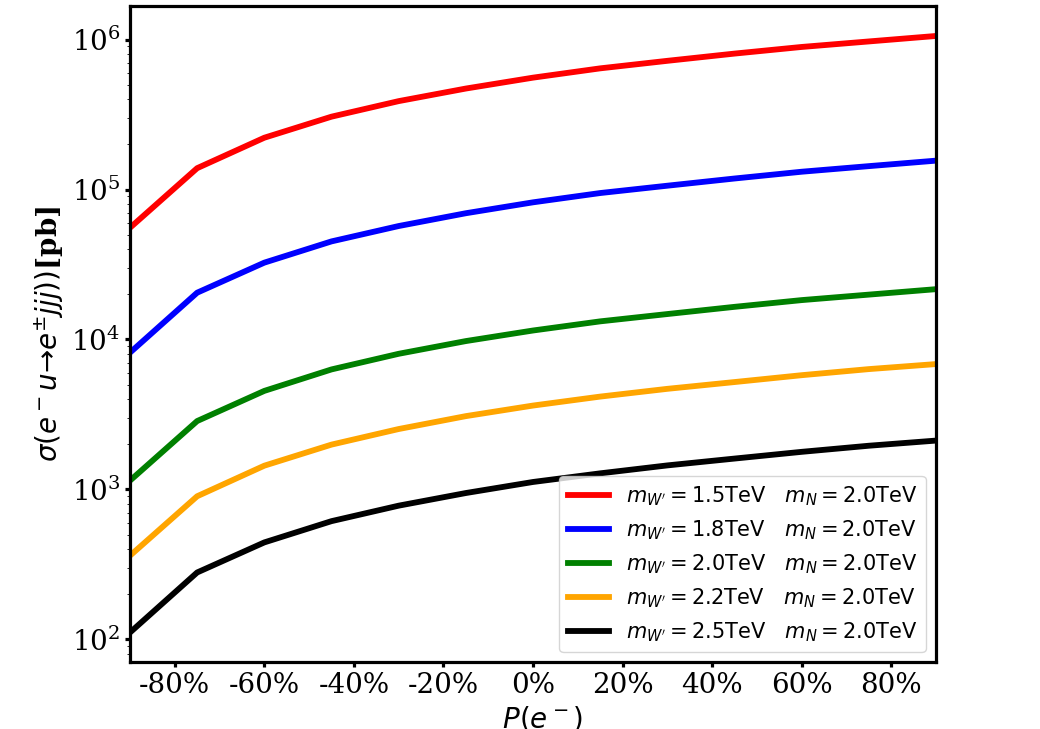}
	}
\captionsetup{justification=raggedright}
	\caption{The distribution of cross section  varies with different values of initial-state electron polarization in the process of $e^ -u \rightarrow e^\pm j j j$ with $\sqrt{s}=5.29~\text{TeV}$.
 }
	\label{fig:1}
\end{figure}

The variation of the cross section with the electron polarization degree is illustrated in Fig.~\ref{fig:1}, where the center-of-mass energy is fixed at $\sqrt{s} = 5.29$ TeV and the polarization degree spans the range from $-90\%$ to $+90\%$. Overall, the scattering cross section increases correspondingly with the rise of the initial-state right-handed polarization degree. Fig.~\ref{fig:1}(a) presents a comparison of five sets of mass parameters ($M_{W^{\prime}}, M_N$), specifically (2.0, 1.5), (2.0, 1.8), (2.0, 2.0), (2.0, 2.2), (2.0, 2.5) TeV. The results indicate that for a fixed $M_{W^{\prime}}$, the cross section increases with $M_N$. In Fig.~\ref{fig:1}(b), it exhibits another group of mass parameters ($M_{W^{\prime}}, M_N$) with values (1.5, 2.0), (1.8, 2.0), (2.0, 2.0), (2.2, 2.0), (2.5, 2.0) TeV. It can be observed that when $M_N$ is a constant, the cross section decreases with the increasing of $M_{W^{\prime}}$.

\begin{figure}[]
	\centering
	\subfloat[]{
		\includegraphics[width=0.47\linewidth]{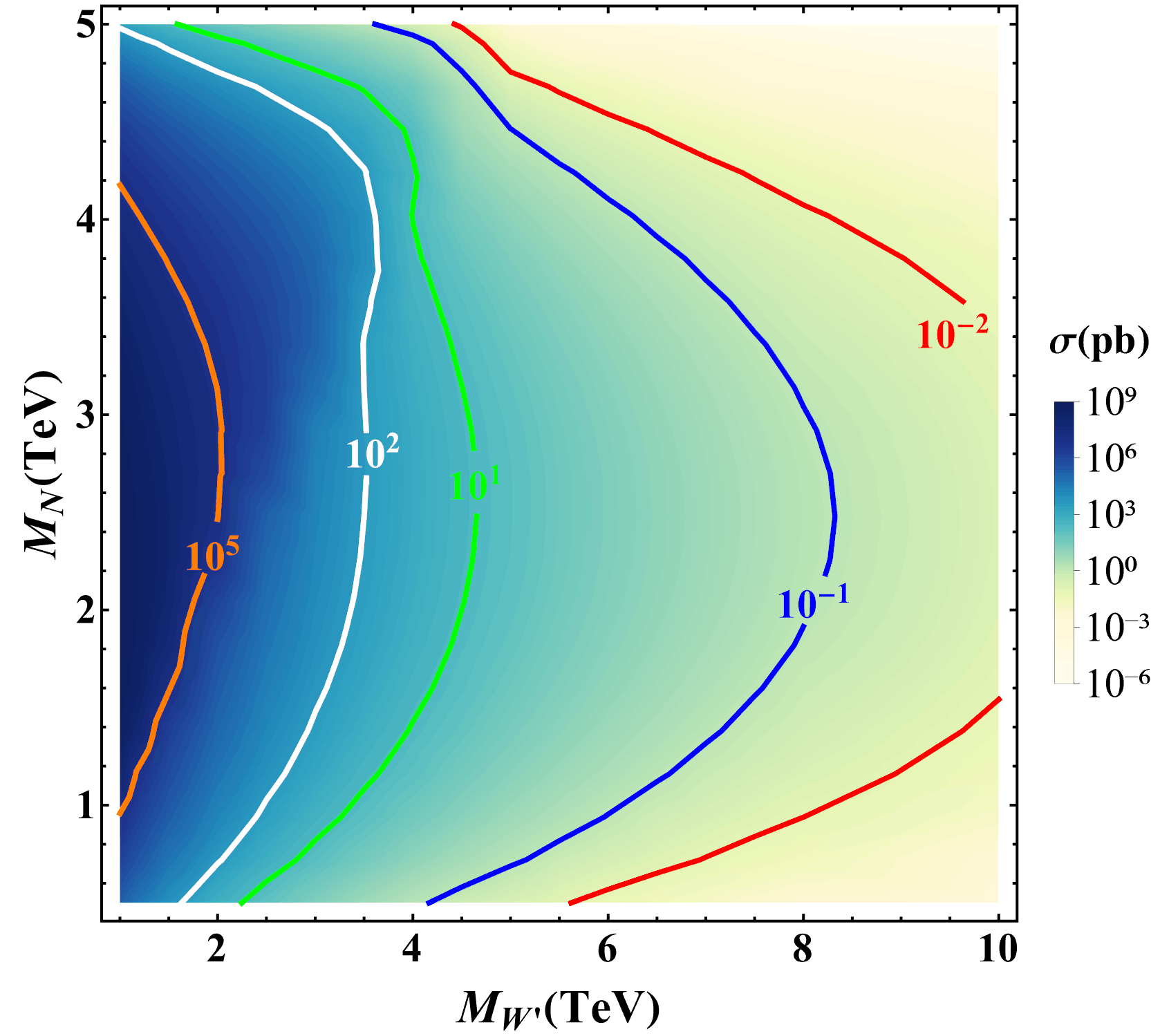}
		\label{1}
	} \hfill
	\subfloat[]{
		\includegraphics[width=0.47\linewidth]{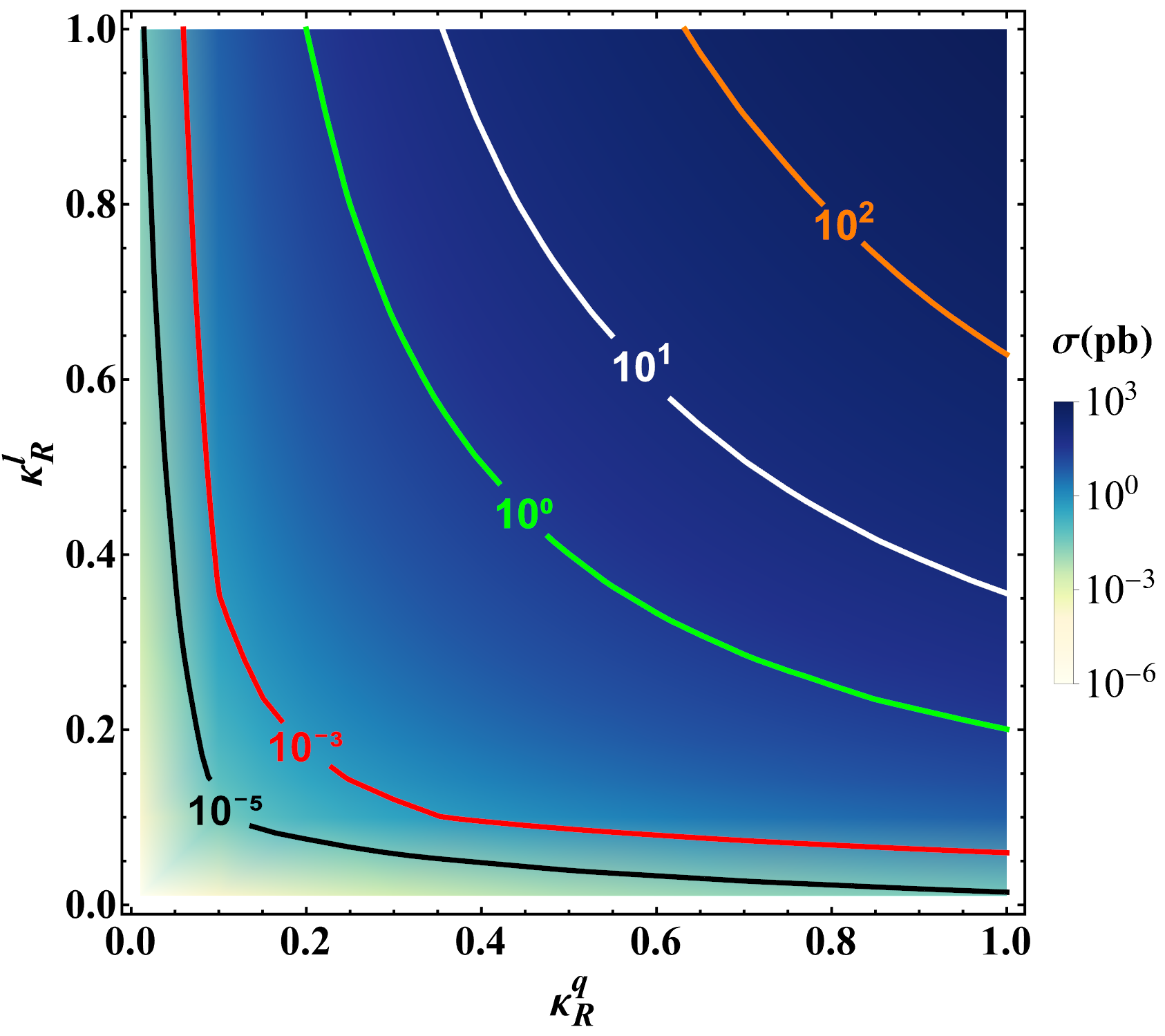}
		\label{2}
	} \\[1ex]
	\subfloat[]{
		\includegraphics[width=0.475\linewidth]{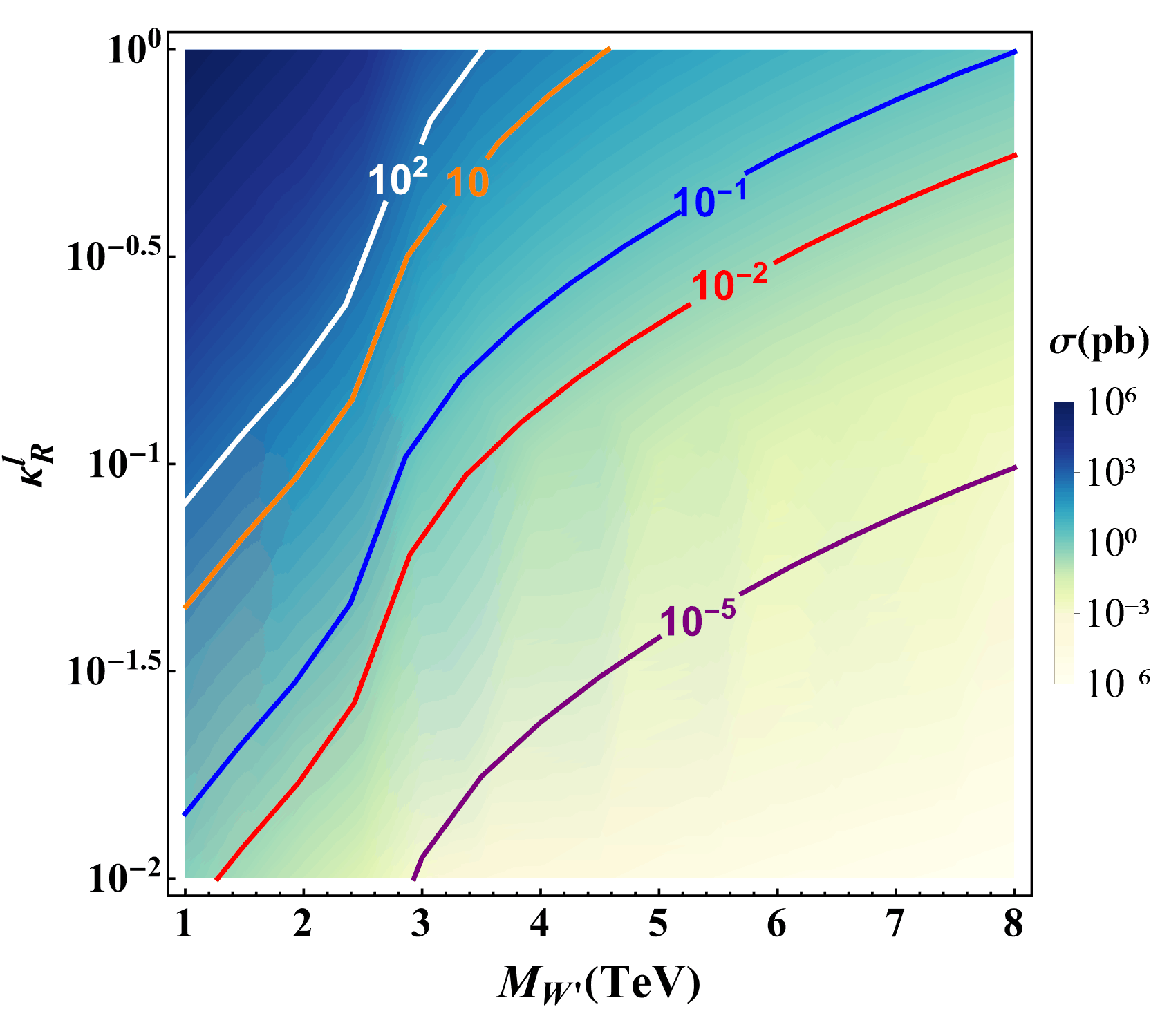}
		\label{3}
	} \hfill
	\subfloat[]{
		\includegraphics[width=0.475\linewidth]{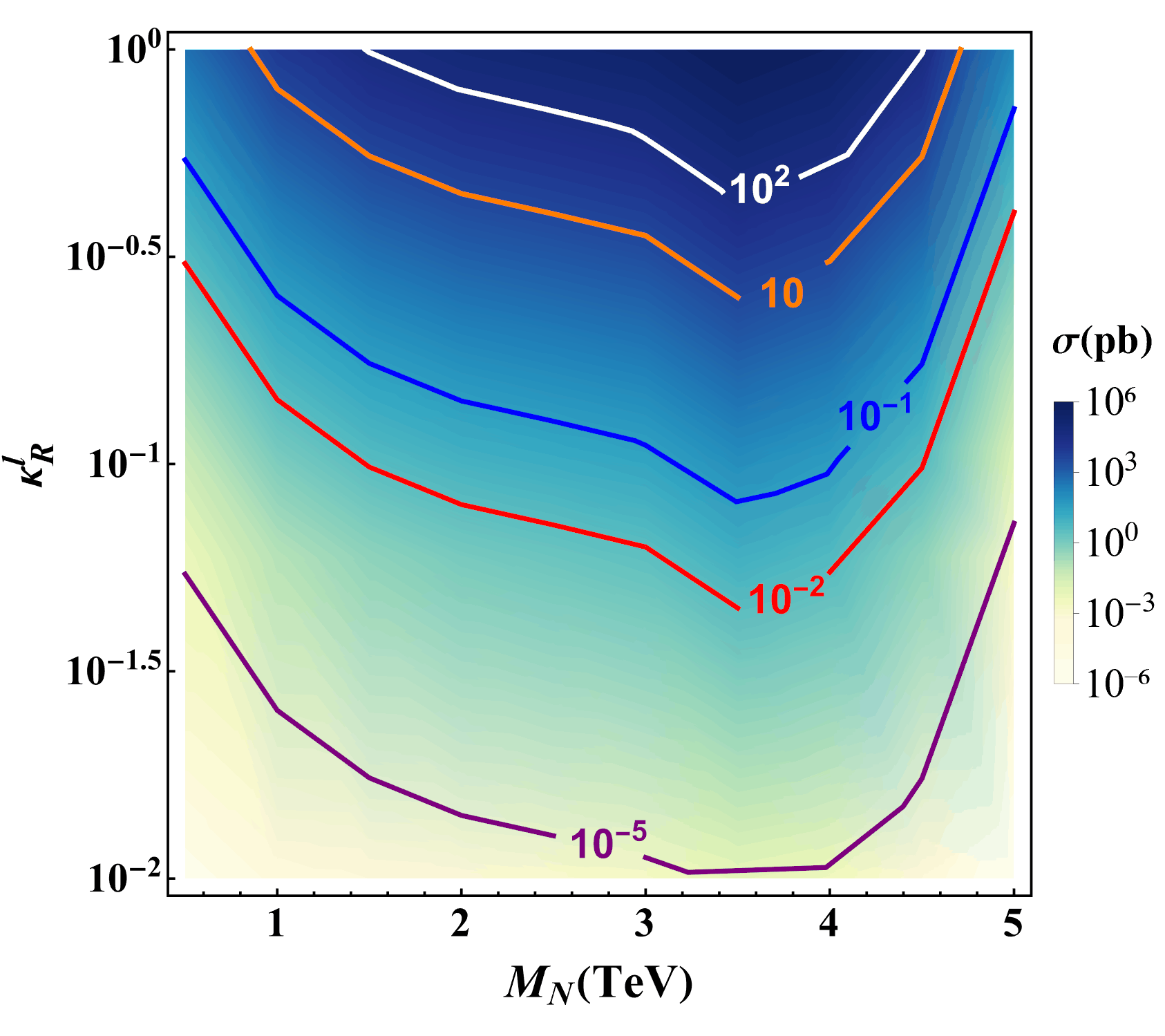}
		\label{4}
	}
\captionsetup{justification=raggedright}
    \caption{Density plots of the distribution of cross section at $\sqrt{s}=5.29~\text{TeV}$ 
as a function of the four parameters $M_{W^{\prime}}$, $M_N$, $\kappa_R^{\ell}$, $\kappa_R^{q}$. (a) versus $M_{W^{\prime}}$ and $M_N$ with $\kappa_R^{\ell}=\kappa_R^{q}=1$;
(b) versus $\kappa_R^{\ell}$ and $\kappa_R^{q}$ with $M_{W^{\prime}}=M_N=3~\text{TeV}$; (c) versus $M_{W^{\prime}}$ and $\kappa_R^{\ell}$ with $M_N=1~\text{TeV}$, $\kappa_R^{q}=1$; 
(d) versus $M_N$ and $\kappa_R^{\ell}$ with $m_{W^{\prime}}=3~\text{TeV}$, $\kappa_R^{q}=1$.}
\label{fig:MCMC_scan}
\end{figure}

In the process of $e^{-}u\to N d \to e^{\pm}jjj$ , the core variables affecting the cross section are fourfold: the mass of $W^{\prime}$ ($M_{W^{\prime}}$), the mass of N ($M_{N}$), the coupling constant to leptons ($\kappa_R^{l}$), and the coupling strength to quarks ($\kappa_R^{q}$). In Fig.~\ref{fig:MCMC_scan}, we present the corresponding two-dimensional density distribution plots, where the multidimensional parameter scan is performed using the Markov Chain Monte Carlo method\cite{Alwall:2014hca}. In Fig.~\ref{fig:MCMC_scan}(a), the density plot displays the distribution of cross section in the $M_{W^{\prime}}$--$M_N$ plane with $\kappa_{R}^{\ell} = \kappa_{R}^{q} = 1$. As the mass of $W^{\prime}$ increases, the cross section exhibits a gradual decline, consistent with the asymptotic behavior of the $t$-channel propagator at fixed mass splitting. In contrast, with increasing $M_N$, the cross section shows a non-monotonic behavior, initially rising and then falling, with its peak occurring in the intermediate $M_N$ region. This feature is attributed to the competition between phase-space volume and decay branching ratios. Typically, when the mass of $W^\prime$ is 10 TeV and the the mass of right-handed neutrino is 3 TeV, the cross section of the studied process still exceeds 10 fb.

In Fig.~\ref{fig:MCMC_scan}(b), when $M_{W'} = M_N = 3\ \text{TeV}$ is fixed, the distribution of cross section varies with the coupling parameters $\kappa_{R}^{\ell}$ and $\kappa_{R}^{q}$. The range of the coupling parameters is $0.01 \leq \kappa_{R}^{\ell},\kappa_{R}^{q} \leq 1$. In the plot that each line represents the cross section size of $10^{-5}, 10^{-3}, 10^0, 10^1, 10^2\ \text{pb}$ respectively. The cross section monotonically increases with the increase of either coupling parameter. When $\kappa_{R}^{\ell}$ and $\kappa_{R}^{q}$ are both close to 0.01, the cross section drops to approximately the $10^{-5}\ \text{pb}$ level. In such a scenario, if it is assumed that more than one signal event is observed, the required integrated luminosity would be 1000 $fb^{-1}$, which remains within the designed luminosity range.

Fig.~\ref{fig:MCMC_scan}(c) and (d) display the influence of the two-dimensional correlation between the mass parameters and the coupling parameters on the cross section. Fig.~\ref{fig:MCMC_scan}(c) shows the distribution of the cross section in the $M_{W'}$–$\kappa_{R}^{\ell}$ plane with $M_N = 1\ \text{TeV}$ and $\kappa_{R}^{q} = 1$. Fig.~\ref{fig:MCMC_scan}(d) shows the distribution of the cross section in the $M_N$–$\kappa_{R}^{\ell}$ plane, with $M_{W'} = 3\ \text{TeV}$ and $\kappa_{R}^{q} = 1$. In Fig.~\ref{fig:MCMC_scan}(c), the contour lines clearly show that the cross section significantly decreases as the mass of $W^{\prime}$ increases, 
while it shows an opposite trend as the coupling constant $\kappa_{R}^{\ell}$ grows. In Fig.~\ref{fig:MCMC_scan}(d), when $\kappa_{R}^{\ell}$ is fixed, the cross section first increases and then decreases with the change of $M_N$, reaching a maximum at a certain intermediate value of $M_N$. This trend occurs because when the mass of the right-handed neutrino exceeds that of the $W^{\prime}$ boson, the decay channel of $N \to W^{\prime} l$ opens up, leading to an enhancement effect on the cross section of the studied process.
From the symmetry of the Feynman diagrams in Fig.~\ref{fig:feynman}, it can be observed that the coupling $\kappa_{R}^{q}$ and $\kappa_{R}^{\ell}$ play identical roles in this process, and their corresponding distributions are similar. Therefore, related plots are not separately presented  in the paper. In summary, this series of density plots systematically reveal the relative importance of the mass and coupling parameters in influencing the cross section in different parameter subspaces.

The angular distribution of final-state particles is an important observable for studying the interactions of $W^{\prime}$ boson. The angular distribution formula typically includes the particle's momentum and polar angle, which describe the particle's motion relative to a reference direction. The expression for the angular distribution is defined as follows:
\begin{equation}
cos \theta=\frac{\boldsymbol{p}_{\boldsymbol{f}}^{*} \cdot \boldsymbol{p}_{\boldsymbol{i}}}{\left|\boldsymbol{p}_{\boldsymbol{f}}^{*}\right| \cdot\left|\boldsymbol{p}_{\boldsymbol{i}}\right|} ~~~,
\end{equation}

\begin{figure}[]
	\centering
	\subfloat[]{
		\includegraphics[width=0.48\linewidth]{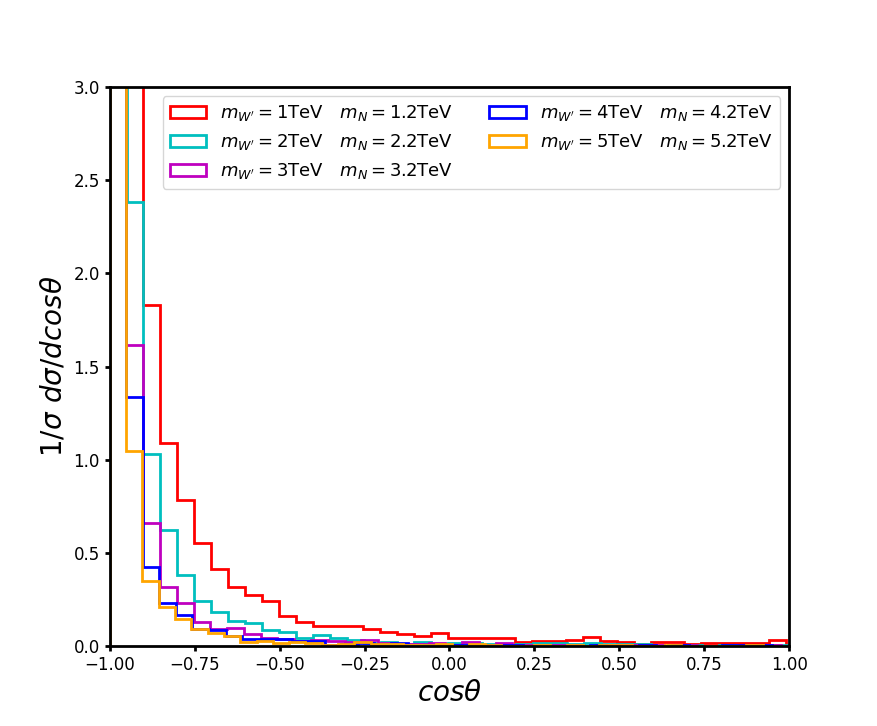}
		\label{cos1}
	} 
	\subfloat[]{
		\includegraphics[width=0.48\linewidth]{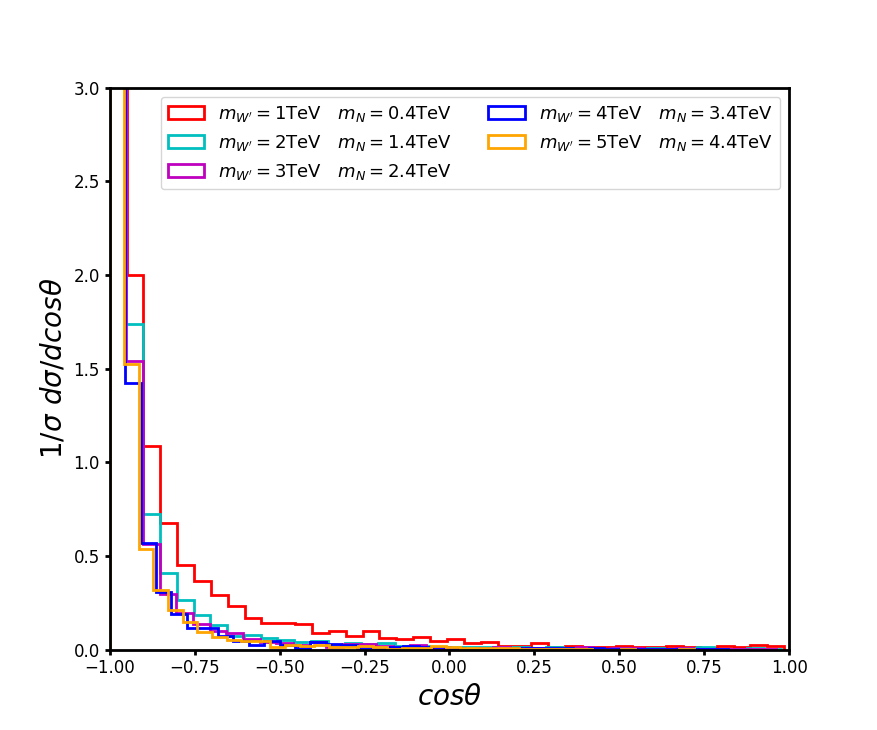}
		\label{cos2}
	}
\captionsetup{justification=raggedright}
	\caption{The angular distribution of the final-state leptons in the $e^{-}p\to e^{\pm}jjj$ process at $\sqrt{s}=5.29$~TeV with different mass splittings.}
	\label{fig:angular_dist}
\end{figure}

 here, $\mathbf{p}_f^*$ and $\mathbf{p}_i$ are the three-momenta of the final and initial particles, respectively. The angular distribution of the final-state electron in the process $e^- u \to e^\pm j j j$ at a center-of-mass energy $\sqrt{s} = 5.29~\text{TeV}$ is shown in Fig.~\ref{fig:angular_dist}, with two distinct fixed mass-splitting conditions between the $W^{\prime}$ boson and the Majorana neutrino $N$. In Fig.~\ref{fig:angular_dist}(a), the mass difference is fixed at $\Delta M = M_{W^{\prime}} - M_N = 600~\text{GeV}$, while in Fig.~\ref{fig:angular_dist}(b), it is fixed at $\Delta M = 200~\text{GeV}$. For each panel, the $W^{\prime}$ mass $M_{W^{\prime}}$ is varied from $1$ to $5~\text{TeV}$. The distributions exhibit a marked forward-backward asymmetry, manifesting as a significant accumulation of events in the large-angle (backward) region ($\cos\theta_{e^-} \to -1$). This pronounced backward peaking is a characteristic signature of the $t$-channel exchange of the heavy $W^{\prime}$ boson. The asymmetry pattern and the normalization of the distributions are observed to depend on both the absolute mass scale $M_{W^{\prime}}$ and the specific mass splitting $\Delta M$, 
 providing distinct kinematic handles for disentangling the underlying new physics parameters.

To better characterize the trend of the final-state particle angular distribution, we define the forward--backward asymmetry as
\begin{equation}
A_{F B}=\frac{\sigma(\cos \theta \geq 0)-\sigma(\cos \theta<0)}{\sigma(\cos \theta \geq 0)+\sigma(\cos \theta<0)}
\end{equation}

\begin{figure}[]
	\centering
\includegraphics[width=0.8\linewidth]{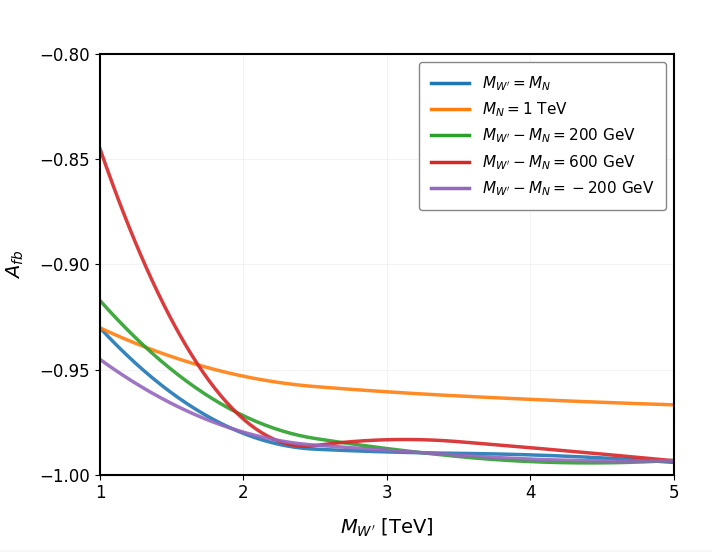}
\captionsetup{justification=raggedright}
	\caption{The forward–backward asymmetry $A_{\text{FB}}$ in the $e^{-}u\to e^{\pm}jjj$ process is shown as a function of the $m_{W^{\prime}}$ boson mass, $\sqrt{s}=5.29~\text{TeV}$.}
	\label{fig:afb}
\end{figure}

Fig.~\ref{fig:afb} shows the variation of the forward-backward asymmetry $A_{\mathrm{FB}}$ for the process  $e^-u \to e^\pm jjj$ with different mass value of $W^{\prime}$ particle. It can be seen from the figure that the absolute value of the asymmetry generally increases with the increase of the $W^{\prime}$ mass, but its specific behavior is significantly affected by the mass splitting  $\Delta M = M_{W^{\prime}} - M_N$ between the $W^{\prime}$ and $N$ in the low $W^{\prime}$ mass region. When $\Delta M = 600$ GeV (red line), the absolute value of the asymmetry $A_{\mathrm{FB}}$ reaches 0.85 at $ M_{W^{\prime}} = 1$ TeV; However, in the higher $W^{\prime}$ mass region, the variation of the asymmetry $A_{\mathrm{FB}}$ becomes more gradual. When the mass of $N$ is set to 1 TeV (yello line), the absolute value of the asymmetry tends to approach 0.95 as the mass of $W^{\prime}$ increases.
A further comparison between $\Delta M =$ 0 (blue line) and the case with a fixed $M_N = 1$ TeV (yello line) reveals that under the condition of mass degeneracy ($M_{W^{\prime}} = M_N$), the absolute value of $A_{\mathrm{FB}}$ continuously increases with the increase of $M_{W^{\prime}}$ and it varies significantly in the low-mass region but less so in the high-mass region. Therefore, the mass splitting has a significant impact on the forward-backward asymmetry and can be used as an important observable to distinguish different mass spectrum structures.

\begin{figure}[]
	\centering
	\subfloat[]{
		\includegraphics[width=0.46\linewidth]{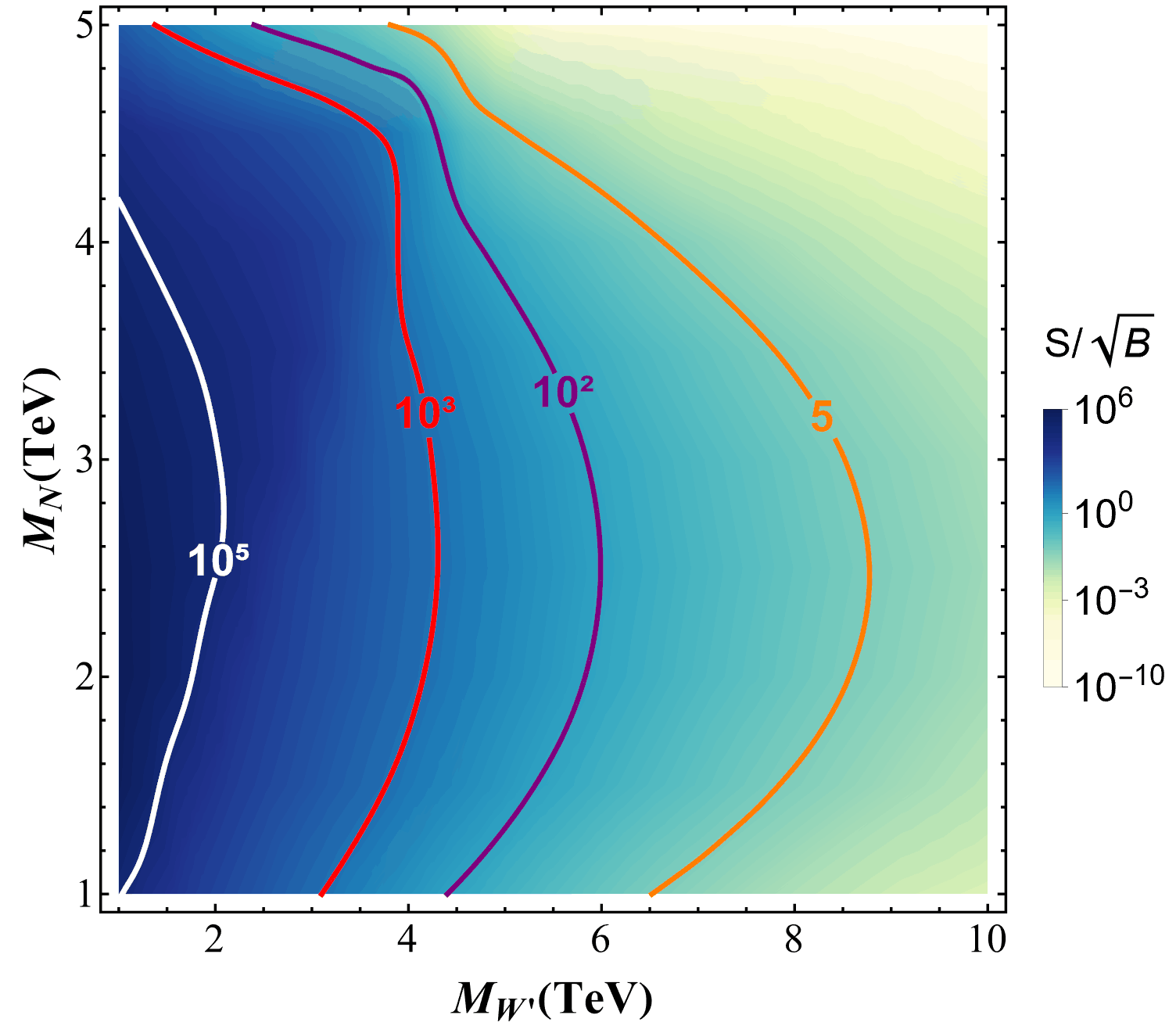}
		\label{sb1}
	} \hfill
	\subfloat[]{
		\includegraphics[width=0.46\linewidth]{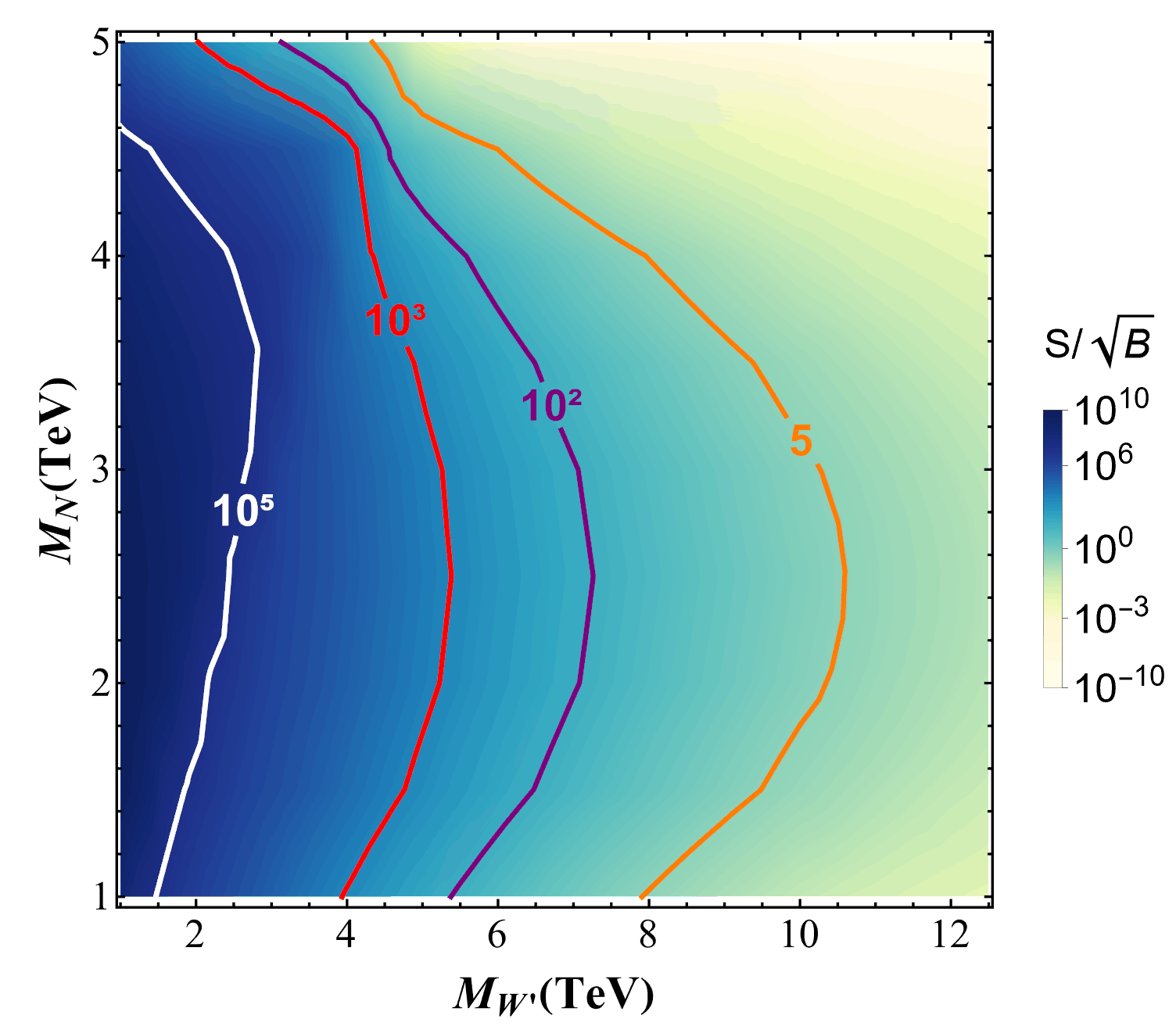}
		\label{sb2}
	} \\[1ex]
	\subfloat[]{
		\includegraphics[width=0.46\linewidth]{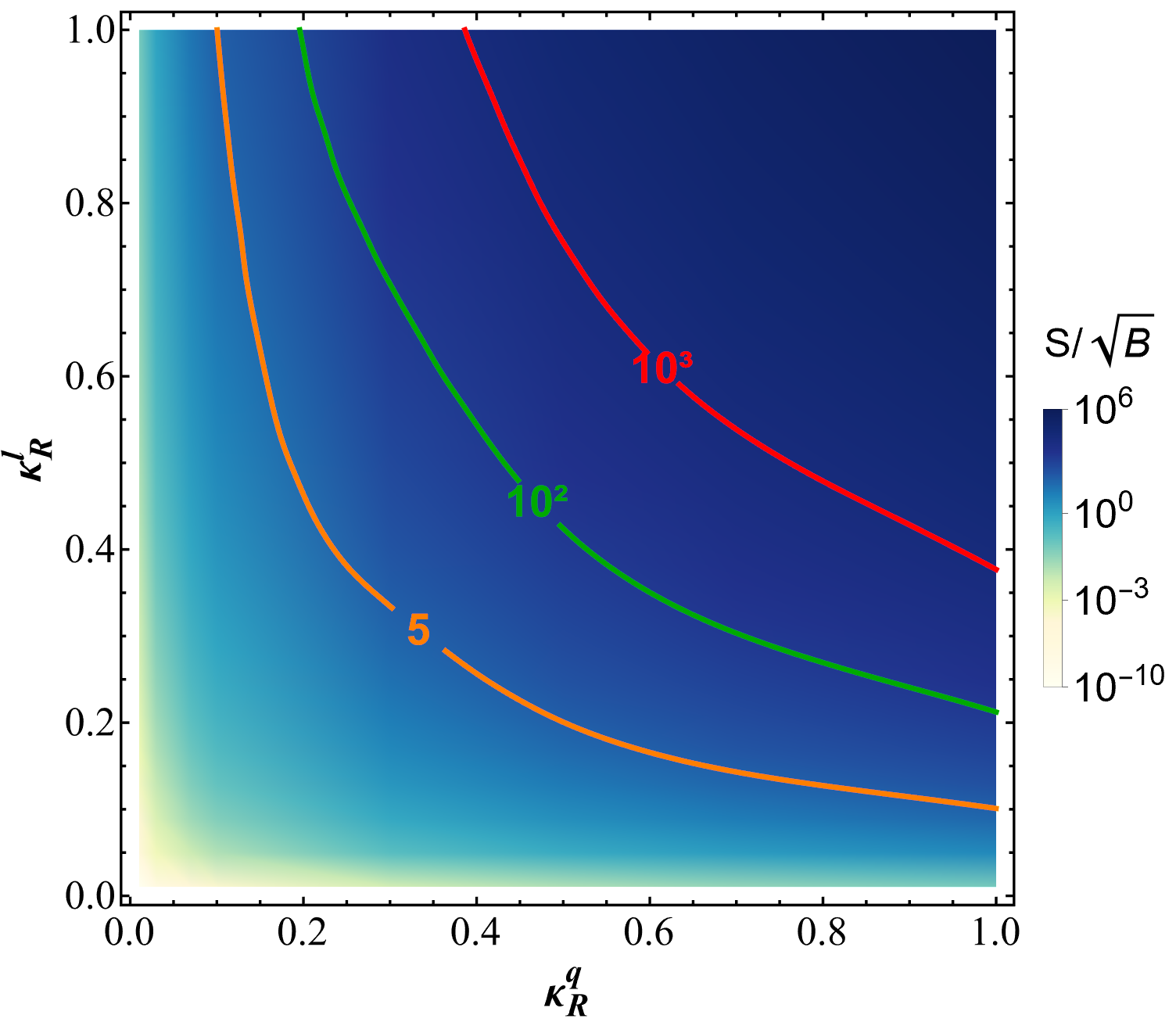}
		\label{sb3}
	} \hfill
	\subfloat[]{
		\includegraphics[width=0.46\linewidth]{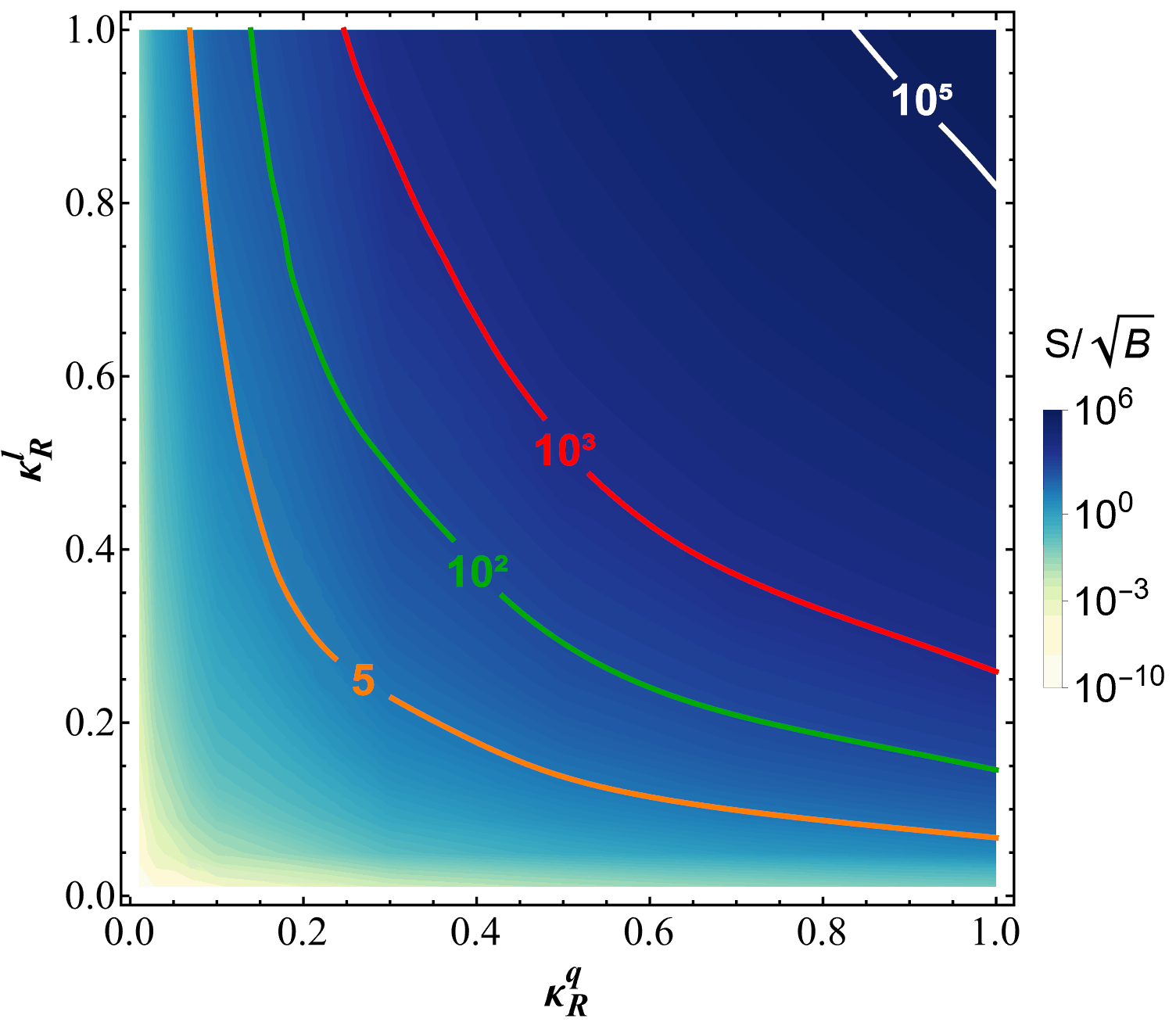}
		\label{sb4}
	}
\captionsetup{justification=raggedright}
	\caption{The density maps of the statistical significance for the $e^{-}u\to e^{-}jjj$ process is presented with $\sqrt{s}=5.29~\text{TeV}$ and the integrated luminosity of  $100~\text{fb}^{-1}$ (left plots) and $2000~\text{fb}^{-1}$ (right plots) respectively. (a) (b)  $M_{W^{\prime}} - M_N$ plane with $\kappa_R^{\ell}=\kappa_R^{q}=1$;
(c) (d)  $\kappa_R^{\ell}- \kappa_R^{q}$ plane with $M_{W^{\prime}}=M_N=3~\text{TeV}$.}
	\label{s/b}
\end{figure}

To further investigate the properties of $W^{\prime}$ at $e^- p$ colliders, we generated the corresponding signal and background processes using  MadGraph5$\_$aMC@NLO\cite{Alwall:2011uj,Alwall:2014hca,Frixione:2021zdp}. Within the Effective Left--Right Symmetric Model we focus on the process $e^{-}u\to e^{-}jjj$ mediated by a $W^{\prime}$ boson, while the background is taken as the Standard-Model process of  
$e^{-}u\to e^{-}jjj$ with $t$-channel $W$ exchange.  Employing the standard definition of statistical significance, we obtain the two-dimensional significance density maps in the $(M_{W^{\prime}}-M_{N})$ plane and $(\kappa_R^{\ell}-\kappa_R^{q})$ plane. The significance $S / \sqrt{B}$ is defined as the ratio of the number of signal events to the square root of  the number of background events. The left and right plots are assumed with the integrated luminosity of  
$100~\text{fb}^{-1}$ and $2000~\text{fb}^{-1}$, respectively. Fig.~\ref{s/b}(a) and (b) illustrate the variation of significance with the masses of $W^{\prime}$ and $N$ when the coupling strength between $W^{\prime}$ and fermions is consistent with that of the Standard Model. As the mass of $W^{\prime}$ increases, the significance gradually decreases, while as the mass of $N$ increases, the significance first rises and then declines. When the mass of $N$ is 2.5 TeV, considering the significance of $5\sigma$ and the integrated luminosity of 100 (2000) $fb^{-1}$, the observable mass range of $W^{\prime}$ can extend up to 9 (11) TeV. When the masses of $W^{\prime}$ and $N$ are fixed at 3 TeV, Fig.~\ref{s/b}(c) and (d) show the variation of significance with the coupling strengths  $\kappa_R^{\ell}$ and $\kappa_R^{q}$. As the coupling strengths increase, the significance also increases significantly. Considering the $5\sigma$ exclusion line, the minimum observable coupling strength is approximately 0.15 times that of the Standard Model.  As shown in Fig.~\ref{fig:MCMC_scan}, if scenarios with larger $W^{\prime}$ masses or mass splittings between $W^{\prime}$ and $N$ are considered, the detectable range of coupling strengths could reach the level of 1\% of the Standard Model.

\section{Summary}\label{sec:4}
The research on additional charged gauge bosons beyond the Standard Model has always been a topic of interest. In this paper, we have thoroughly investigated the interactions of the $W^{\prime}$ particle at electron-proton colliders such as the  LHeC and the FCC-he. We first studied the simplest $e^- u \rightarrow \nu_e d$ process mediated by $t$-channel $W^{\prime}$ exchange within  the $W^{\prime}$ Effective Model. This channel serves as a direct probe of $W^{\prime}$ and fermions interactions, and our results indicate that the cross section decreases as $M_{W^{\prime}}$ increases. Within the mass range $M_{W^{\prime}}$=1$\sim$ 5 TeV, the cross section varies roughly between $10^{-1} \sim 10^{-4}$pb. When the electron is polarized, the cross section increases with the electron polarization degree $P(e^-)$. With $M_{W^{\prime}} = 1~\text{TeV}$ and $P(e^-)=80\%$, the cross section is 0.87 pb. 

The Effective Left-Right Symmetric Model is a more promising new physics model for explaining neutrino mass and CP violation, among other issues. However, this theoretical framework does not encompass the $e^- u \rightarrow \nu_e d$ process. As a result, we have investigated the $e^- u \rightarrow e^\pm jjj$ process.
 This process involves $t$-channel $W^{\prime}$ exchange and can provide sensitivity to the $W^{\prime}$ mass and mass spectrum of heavy neutrinos through an intermediate Majorana neutrino ($N$). A detailed scan of the parameter space ($M_{W^{\prime}}$, $M_N$, $\kappa_R^{\ell}$, $\kappa_R^{q}$) revealed rich phenomenology. With $\sqrt{s} = 5.29$ TeV, the cross section is $4.63\times10^{-4}$ pb with 
 $M_{W^{\prime}}=M_N=5$ TeV. Meanwhile, the size of the mass splitting between $W^{\prime}$ and $N$ also affects the distribution of cross section. Under specific mass splitting parameters between $W^{\prime}$ and $N$, the cross section of $e^- u \rightarrow e^\pm jjj$ can significantly increase. The polarization of the initial-state electrons has a significant impact on the cross section. When the electron polarization reaches $90\%$, the cross section can be twice as large as that in the unpolarized scenario. Additionally, the forward-backward asymmetry $A_{\mathrm{FB}}$ shows a pronounced backward peak asymmetry, characterized by a significant accumulation of events in the large-angle (backward) region. After considering the backgrounds related to the Standard Model, we have provided the signal significance across different parameter spaces. When $\sqrt{s} = 5.29$ TeV, the upper limit of detectable $M_{W^{\prime}}$ is approximately 9 (11) TeV with $\mathcal{L} = 100 ~(2000)$ $fb^{-1}$. It should be particularly noted that the $e^- u \rightarrow e^+ jjj$ process has no corresponding processes within the Standard Model, which means that if related events are observed, they would serve as important evidence for new physics beyond the Standard Model. In summary, the proposed electron-proton collider, especially the high-energy, high-luminosity FCC-he, provides a powerful and complementary environment for searching for $W^{\prime}$ bosons at the TeV scale. The study of exotic gauge bosons at $e^-p$ colliders
  offers a new perspective for the search of new physics.

\section*{Acknowledgement}
This work is supported by  National Natural Science Foundation of China under Grant No.12505112, Natural Science Foundation of Shandong Province (ZR2024QA138, ZR2022M \allowbreak A065), State Key Laboratory of Dark Matter Physics, and University of Jinan Disciplinary Cross-Convergence Construction Project 2024 (XKJC-202404).

\bibliographystyle{apsrev4-2}
\bibliography{references}

\end{document}